\documentclass
[secnumarabic,amsmath,amssymb,showpacs,nobibnotes,aps,pre,12pt]{revtex4-1}
\usepackage{amssymb}

\usepackage{amssymb}

\usepackage{graphicx}
\usepackage{dcolumn}
\usepackage{bm}
\usepackage{color}

\begin{document}
\title{From Liouville's theorem to quasilinear, nonlinear stochastic, and fractional transport, a multi-scale kinetics of plasma turbulence}

\author{Shaojie Wang}
\email{wangsj@ustc.edu.cn}
\affiliation{Department of Modern Physics, University of Science and Technology of China, Hefei, 230026, China}
\date{\today}

\begin{abstract}
From Liouville's equation, a phase-space multi-scale transport equation is systematically derived. The proposed phase-space multi-scale transport equation based on the first principle indicates that the nonlinear stochastic transport is due to the micro-turbulence scattering while the familiar quasilinear transport is due to the long-range correlation of the meso-scale coherent modes; and more interestingly, it demonstrates a systematic derivation of the fractional transport equation that may provide a new view-angle to understand the anomalous transport observed in plasma turbulence. The multi-scale kinetics suggests a new approach to confinement improvement through the formation of nonlinearly self-organized stable large-scale structures.
\end{abstract}


\maketitle

\section{\label{sec1}INTRODUCTION}
Transport is one of the most challenging scientific problems for many years\cite{ChandrasekharRMP43,ZaslavskyPR02}. Whereas the motion of a single particle is described by the Hamiltonian dynamics\cite{CaryPR81,CaryAoP83,LittlejohnJMP82,BrizardPLA01}, the physical kinetics of a many-particle system is governed by Liouville's (Vlasov's) equation\cite{HazeltineBook98}. In magnetic confinement fusion researches, the problem of transport is treated by solving Liouville's equation for the particle distribution in the phase space, and the kinetic equation is nonlinearly coupled with Maxwell's equations, because the motion of charged particles depends on the electromagnetic fields that depend on the distribution of charged particles \cite{HazeltineBook98, KadomtsevBook70, BalescuBook05, DiamondBook10}. Generally, the fields can be decomposed into the fluctuating part and the averaged part \cite{HazeltineBook98}. The fluctuations due to the particle discreteness and the collective instabilities of the plasma are known respectively as the collisional dissipation described by the Fokker-Planck (FP) equation\cite{HintonRMP76} and the micro-turbulence\cite{KadomtsevBook70, BalescuBook05, DiamondBook10,ConnorPPCF94} described by the gyrokinetic Vlasov (Liouville) equation\cite{FriemanPF82,BrizardRMP07} which does not explicitly display the nonlinear stochastic dissipation.
The standard FP transport equation\cite{ChandrasekharRMP43,EscandePRL07,BalescuBook05,DiamondBook10,ConnorPPCF94} describes the transport fluxes in terms of the local thermal dynamic forces. However, the anomalous transport is still a difficult unsolved problem\cite{ConnorPPCF94}. In fact, there are counter-examples to the local transport paradigm\cite{GentlePoP95}, and the fractional kinetics has also been investigated to understand the anomalous transport\cite{ZaslavskyPR02}.
Recently, Liouville's equation has been solved in limiting cases to explicitly display the nonlinear turbulence dissipation in terms of the FP operator\cite{WangPRE13,WangPoP13} and the modified quasilinear diffusion due to the coherent waves\cite{KominisPRL10,WangPoP12}. However, it is not clear what is the physical difference between the quasilinear transport and the nonlinear stochastic transport, and how can one derive a fractional transport equation from Liouville's equation.

In this paper, we shall derive a transport equation from Liouville's theorem to describe the multi-scale transport in phase-space, which includes the quasilinear, the nonlinear stochastic, and the fractional transport. Although the purpose of this work is to provide a new framework in first principle to understand the anomalous transport in fusion plasmas due to the micro-turbulence and coherent waves, the methods developed are also useful in astro-physics\cite{ChandrasekharRMP43}, space plasma turbulence\cite{ChenJGR99}, and statistical physics of nonlinear and chaotic systems\cite{HazeltineBook98,ZaslavskyPR02}. 

The remaining parts of this paper is organized as follows. In Sec. II, we decouple the irregular motion from the regular motion by using the Lie-transform method. In Sec. III, we derive the multi-scale phase space transport equation. In Sec. IV, we derive the nonlinear stochastic transport and quasilinear transport equation. In Sec. V, we derive the fractional transport equation. In Sec. VI, we summarize the main results and discuss the implications of the multi-scale kinetics to the improvement of confinement.

\section{\label{sec2}DECOUPLING THE IRREGULAR FROM THE REGULAR MOTION}
We begin with the single particle Hamiltonian dynamics. Following Brizard's geometrical view\cite{BrizardPLA01} on the Lie-transform perturbation method\cite{CaryPR81,CaryAoP83,LittlejohnJMP82}, we write down the fundamental one-form (Lagrangian),
\begin{equation}
 d\mathcal{A} = \Gamma_i dZ^i-\left(H_0+\epsilon H_1\right)dt-Sd{\epsilon},\label{eq:VP}
\end{equation}
where $\bm Z$ is the 6-dimensional phase-space noncanonical coordinates. The Hamiltonian is split into the unperturbed part $H_0$ and the perturbation part $H_1$, with $\epsilon$ a formal perturbation parameter. Note that a general electro-magnetic perturbation problem can be described by Eq. (\ref{eq:VP})\cite{WangPRE13,WangPoP13}.

The particle motion is governed by the unperturbed Poisson brackets and the perturbed Hamiltonian,
\begin{subequations}
\begin{eqnarray}
\partial_t         Z^i\left(t,\epsilon\right)&=&\left\{Z^i,H_0+\epsilon H_1\right\},\label{eq:t-dynamics}\\
\partial_{\epsilon}Z^i\left(t,\epsilon\right)&=&\left\{Z^i,S\right\}.\label{eq:e-dynamics}
\end{eqnarray}
\end{subequations}
Eqs. (\ref{eq:t-dynamics} and \ref{eq:e-dynamics}) describe the $t-$ dynamics and the $\epsilon -$ dynamics, respectively, which can be understood by examining $ d Z^i\left(t,\epsilon\right)=\left[ dt\partial_t +d{\epsilon}\partial_{\epsilon}\right]Z^i\left(t,\epsilon\right)$,
and keeping in mind that $t$ and $\epsilon$ are two parameters independent of each other\cite{BrizardPLA01}.

The Lie-transform generating function $S$ for the $\epsilon -$ flow is given by
\begin{equation}
 \partial_t S+\left\{S,H_0+\epsilon H_1\right\}=\partial_{\epsilon}H=H_1.\label{eq:S}
\end{equation}

Eqs. (\ref{eq:t-dynamics} and \ref{eq:e-dynamics}) indicate that the particle orbit from $t_0$ to $t$ can be found by first pushing the particle over $t$ with $\epsilon =0$ held and then pushing the particle over $\epsilon$ with $S\left(t_0\right)=0$. The solution of Eq. (\ref{eq:e-dynamics}) is given by\cite{BrizardPLA01}
\begin{equation}
Z^i\left(t,\epsilon\right)=\mathcal T_{\epsilon}^{-1} Z^i\left(t,t_0\right) \equiv exp\left[-\int_{0}^{\epsilon}d{\epsilon}\left\{S\left(\bm Z,t,\epsilon\right),\right\}\right]  Z^i\left(t,0\right).\label{eq:ZeZ0}
\end{equation}
Note that the familiar solution to Eq. (\ref{eq:t-dynamics}) with $\epsilon =0$ is
\begin{equation}
Z^i\left(t,0\right)=\mathcal T_{t,0}^{-1} Z^i\left(t,0\right) \equiv exp\left[-\int_{t_0}^{t}d{t}\left\{H_0,\right\}\right] Z^i\left(t_0,0\right).\label{eq:Zt0}
\end{equation}
Since Poisson's brackets are independent of $\epsilon$, one finds
\begin{equation}
\mathcal T_{\epsilon}^{-1} = exp\left[-\left\{\int_0^{\epsilon}d{\epsilon}S\left(\bm Z,t,\epsilon\right),\right\}\right] \equiv exp\left[-\left\{S_{\int},\right\}\right].\label{eq:Te}
\end{equation}

Let
\begin{equation}
G^i  = \left\{Z^i,S_{\int}\right\}.\label{eq:G}
\end{equation}
One finds
\begin{equation}
 \mathcal T_{\epsilon}^{-1} = exp\left[\bm G \cdot \partial_{\bm Z}\right]= exp\left[G^i \partial_i\right], \label{eq:TeG}
\end{equation}
and the phase-space incompressibility condition\cite{WangPoP12,WangPRE13,WangPoP13,BrizardRMP07},
\begin{equation}
\partial_{\bm Z}\cdot \bm G \equiv \frac{1}{\mathcal J}\partial_i\left(\mathcal J G^i\right) =0,\label{eq:divG}
\end{equation}
with $\mathcal J$ the Jacobian of the phase-space.

The generating function can be solved in a perturbative way. Expanding $S$ as
\begin{equation}
S\left(\bm Z,t,{\epsilon}\right) =\sum_{n=1}^\infty \epsilon ^{n-1} S_n\left(\bm Z,t\right),
\end{equation}
and substituting into Eq. (\ref{eq:S}), one finds
\begin{subequations}
\begin{eqnarray}
\partial_t S_1+\left\{S_1,H_0\right\}&=&H_1,\\
\partial_t S_n+\left\{S_n,H_0\right\}&=&-\left\{S_{n-1},H_1\right\},n\geq 2;
\end{eqnarray}
\end{subequations}
\begin{equation}
S_{\int}\left(\bm Z,t,{\epsilon}\right) =\sum_{n=1}^\infty \epsilon ^n \frac{1}{n}S_n\left(\bm Z,t\right). \label{eq:S*}
\end{equation}
The above perturbative solution is slightly different from Dragt-Finn's scheme\cite{CaryPR81}, however, it exactly agrees with Dragt-Finn's result up to the second order. The convergence of the above scheme can be understood by examining a unperturbed system, with a time-independent Hamiltonian $H$. If one writes the Hamiltonian formally as $\epsilon H$, and runs the above Lie-transform calculation, one finds the exact solution.

We note that with the above transform, the irregular motion described by $H_1$ is decoupled from the regular motion described by $H_0$.

\section{\label{sec3}MULTI-SCALE TRANSPORT IN PHASE SPACE}
The particle distribution function $f\left(\bm Z,t,\epsilon\right)$ satisfies Liouville's theorem, $df=0$, with
\begin{equation}
d f= dt\left[\partial_t +\partial_tZ^i\left(t,\epsilon\right)\partial_i\right]f +d{\epsilon}\left[\partial_{\epsilon} +\partial_{\epsilon}Z^i\left(t,\epsilon\right)\partial_i\right]f,\label{eq:f-te}
\end{equation}
where again $t$ and $\epsilon$ are understood as independent of each other.
The $t-$ kinetics is given by
\begin{subequations}
\begin{eqnarray}
0&=&\partial_t f\left(\bm Z,t,\epsilon\right)+\partial_t Z^i\left(t,\epsilon\right)\partial_i f ,\label{eq:f-ta}\\
&=&\partial_t f+\left\{f,H_0+\epsilon H_1\right\},\label{eq:f-tb}
\end{eqnarray}
\end{subequations}
and the $\epsilon -$ kinetics is given by
\begin{subequations}
\begin{eqnarray}
0&=&\partial_{\epsilon} f\left(\bm Z,t,\epsilon\right)+\partial_{\epsilon} Z^i\left(t,\epsilon\right)\partial_i f ,\label{eq:f-ea}\\
&=&\partial_{\epsilon} f+\left\{f,S\right\}. \label{eq:f-eb}
\end{eqnarray}
\end{subequations}

Given the distribution at $t_0$, $f\left(\bm Z,t_0,\epsilon\right)$, one can find the distribution at $t$ by first integrating Eq. (\ref{eq:f-tb}) with $\epsilon=0$ held (along the unperturbed orbit),
\begin{equation}
0=\partial_t f\left(\bm Z,t,0\right)+\left\{f,H_0\right\},\label{eq:f-t0}
\end{equation}
to find
\begin{equation}
 f\left(\bm Z,t,0\right)=\mathcal{T}_{t,0}^{-1}f\left(\bm Z,t_0,\epsilon\right),\label{eq:Tev}
\end{equation}
and then integrating Eq. (\ref{eq:f-eb}) over $\epsilon$ with $S\left(t_0\right)=0$,
\begin{equation}
f\left(\bm Z,t,\epsilon\right)-f\left(\bm Z,t,0\right)=\left[\mathcal {T}_{\epsilon}^{-1} -1\right]f\left(\bm Z,t,0\right)\equiv\mathcal {C}_{\int} f\left(\bm Z,t,0\right).\label{eq:f-pb}
\end{equation}

The nonlinear turbulence scattering operator $\mathcal C_{\int}$ can be expanded by using Eq. (\ref{eq:TeG}) and Eq. (\ref{eq:divG}),
\begin{equation}
\mathcal C _{\int}=\partial_{\bm Z}\cdot \sum_{n=0}^\infty   \bm G \left(\bm G \cdot \partial_{\bm Z}  \partial_{\bm Z}\cdot  \bm G \right)^n\left[\frac{1}{(2n+1)!} +\frac{1}{(2n+2)!}\bm G\cdot \partial_{\bm Z}\right],\label{eq:f-pbex}
\end{equation}

Setting $t_0=t-\tau$, one can use Eq. (\ref{eq:Tev}) and Eq. (\ref{eq:f-pb}) to evolve the distribution over a time interval $\tau$.
Eqs.(\ref{eq:Tev}-\ref{eq:f-pbex}) give the general solution to the Liouville (Vlasov) equation, including the complexity of multi-scale or nonlocal kinetics, which will be clarified in the following.

Note that the above Lie-transform solution can be understood by the usual characteristic method.
Let $\Delta \bm Z\left(t\right)=\mathcal {T}_{\epsilon}^{-1} \bm Z\left(t,0\right)-\bm Z\left(t,0\right)$, then $\bm Z\left(t,\epsilon\right)=\bm Z\left(t,0\right)+\Delta \bm Z\left(t\right)$. The standard characteristic method gives the solution
\begin{equation}
f\left[\bm Z,t,\epsilon\right]=f\left(\bm Z-\Delta \bm Z,t,0\right).\label{eq:char}
\end{equation}
Writing $\Delta \bm Z\left(t\right)$ in terms of the Lie-transform expansion [$-\Delta \bm Z\left(t\right)=\bm G +\frac{1}{2}\bm G \cdot \partial_{\bm Z} \bm G+...$], and making the Taylor expansion of Eq. (\ref{eq:char}), one finds the solution by the characteristic method exactly agrees with the Lie-transform method; the point is that Eq. (\ref{eq:char}) is the scalar invariance rule of the pull-back transform\cite{CaryAoP83}[c. f. Eq. (\ref{eq:f-pb}) and Eq. (\ref{eq:f-pbex})].

Truncating Eq. (\ref{eq:f-pbex}) to the second order results to
\begin{equation}
\mathcal C _{\int}= \partial_{\bm Z}\cdot\left[\bm G +\frac{1}{2}\bm G \bm G\cdot \partial_{\bm Z}\right],\label{eq:f-pb2}
\end{equation}
which corresponds to the local FP transport theory\cite{KominisPRL10,WangPoP12,WangPRE13,WangPoP13}], when $\bm G$ is computed by using the second order expansion of Eq. (\ref{eq:S*}).

However, the truncation scheme should be carefully examined. The second-order truncation, Eq. (\ref{eq:f-pb2}), requires $G^i\partial_i f\ll f$,
while the second-order truncation in Eq. (\ref{eq:S*}) requires that the real orbit does not deviate much away from the unperturbed orbit within $\tau$. This is clearly valid for the quasilinear transport\cite{KominisPRL10,WangPoP12}. In the nonlinear stage, one can keep $\tau$ short enough to make the second-order expansion valid, and the above scheme can be repeatedly used to advance the distribution over time; if this short $\tau$ is comparable to the turbulence correlation time, the local diffusive transport can be simply revealed\cite{WangPRE13,WangPoP13}; the "$\tau$ constraint" has been discussed in detail in Ref. [\onlinecite{WangPoP13}].

To proceed, we split the distribution into the fluctuating part $\tilde f$ and the ensemble-averaged part:  $f=\tilde f +F$, $F\equiv \langle f\rangle$.
In a usual stochastic system\cite{ChandrasekharRMP43}, scales may be separated by the typical correlation time $\tau_c$ and the typical correlation length $\lambda _c$, which is defined as follows. When $\tau \geq\tau_c$, $\bm G \bm G /\tau \sim\lambda_c^2/\tau_c$. The scale-length of the ensemble-averaged distribution $F$ is denoted by $l$, with $l\gg \lambda_c$. Let $\delta \equiv \lambda_c /l\ll 1$. Then the averaged distribution evolves on the time scale $\delta ^{-2}\tau_c$. The typical scale-length of the fluctuating part of distribution $\tilde f$ is $ \sim \lambda_c$ and its time scale is $\sim \tau_c$. If one determines to examine the micro-scale kinetics of the fluctuation, the down limit of time interval can approach zero; if one investigates the large-scale kinetics, the down limit of time interval is $\tau_c$ \cite{ChandrasekharRMP43}. The ensemble average is defined as
\begin{equation}
\left\langle f\left(\bm Z\right)\right\rangle=Lim_{l\gg\Delta Z\geq \lambda_c} \frac{1}{\Delta \bm Z ^6}\int_{\Delta \bm Z ^6}d \bm Z ^6 f\left(\bm Z\right). \label{eq:ea}
\end{equation}
The ensemble average over the phase-space volume element with the scale length larger than the turbulence correlation length essentially smooths out the stochastic fluctuations. Obviously, $\Delta \bm Z ^6$ is similar to the usual fluid element; it is infinitely large in the microscopic view, but infinitely small in the macroscopic view.

It follows from Eq. (\ref{eq:Tev}) and Eq. (\ref{eq:f-pb}) that the evolution of the fluctuating part and the ensemble-averaged part of distribution can be written as
\begin{subequations}
\begin{eqnarray}
\tilde f\left(\bm Z,t,0\right)&=&\mathcal{T}_{t,0}^{-1} \tilde f\left(\bm Z,t-\tau,\epsilon\right),\label{eq:Tev1}\\
F\left(\bm Z,t,0\right)&=&\mathcal{T}_{t,0}^{-1}F\left(\bm Z,t-\tau,\epsilon\right);\label{eq:TevF}
\end{eqnarray}
\end{subequations}
\begin{subequations}
\begin{eqnarray}
&& \tilde f\left(\bm Z,t,\epsilon\right)-\tilde f\left(\bm Z,t,0\right)=\mathcal {C}_{\int}  \tilde f\left(\bm Z,t,0\right) + \tilde {\mathcal {C}_{\int}} F\left(\bm Z,t-\tau,\epsilon\right)-\mathcal {O}\left[\langle \mathcal {C}_{\int} \tilde f \rangle\right],\label{eq:pb1}\\
&& F\left(\bm Z,t,\epsilon\right)-F\left(\bm Z,t-\tau,\epsilon\right)=\langle \mathcal {C} _{\int}\rangle F\left(\bm Z,t-\tau,\epsilon\right)+\mathcal {O}\left[\langle \mathcal {C} _{\int} \tilde f \rangle\right],\label{eq:pbF}
\end{eqnarray}
\end{subequations}
where we have assumed that $\langle \mathcal {C}_{\int} \tilde f \rangle=0$, which should be justified for $\tau$ longer than the turbulence correlation time $\tau_c$ in a stochastic turbulence.

If one sets $\tau > \tau_c$, one has $\bm G >\lambda_c$. Clearly the second-order expansion of Eq. (\ref{eq:pb1}) breaks down for the case of strong turbulence due to the slow convergence of $\mathcal {C}_{\int} \tilde f$ [ c. f., Eq. (\ref{eq:f-pbex})]. The combination of Eq. (\ref{eq:pb1}), Eq. (\ref{eq:f-pbex}) and Eq. (\ref{eq:S*}) demonstrates the complexity of the hierarchy folding of the orbits and the mode-mode coupling of the micro-kinetics; when looking at the asymptotic limit $t\rightarrow\infty$, the folding goes to infinity. By examining the characteristic method, Eq. (\ref{eq:char}), one understands that Eq. (\ref{eq:pb1}) includes nonlocal behaviors in micro-kinetics, since $\Delta \bm Z \left(\tau_c\right) \sim \bm G \left(\tau_c\right)$ may be larger than the typical wavelength of the micro-kinetics. Identifying this complexity of the nonlocal behavior and the multi-scale hierarchy of the micro-kinetics may help one to understand the random-phase of the microscopic fluctuations\cite{BalescuBook05}.
In the practical computation where the resolution of the micro-scale kinetics is needed, the convenience of the second-order truncation should be restricted under the condition $\Delta t = \tau \sim \delta \tau_c$.

However, the large-scale kinetics is given by Eq. (\ref{eq:pbF}), with $\tau \geq \tau_c$, and
\begin{equation}
\left\langle \mathcal C_{\int}\right\rangle= \partial_{\bm Z}\cdot \sum_{n=0}^\infty \left[\frac{\left\langle  \bm G \left(\bm G \cdot \partial_{\bm Z}  \partial_{\bm Z}\cdot  \bm G \right)^n\right\rangle}{(2n+1)!} +\frac{\left\langle\bm G \left(\bm G \cdot \partial_{\bm Z}  \partial_{\bm Z}\cdot  \bm G \right)^n\bm G\right\rangle}{(2n+2)!}\cdot\partial_{\bm Z}\right] . \label{eq:nlt}
\end{equation}
which accounts for a multi-scale or nonlocal transport.

\section{\label{sec4}FOKKER-PLANCK EQUATION, NONLINEAR STOCHASTIC TRANSPORT AND QUASILINEAR TRANSPORT}

The FP transport equation is simply the second-order truncation written in the limit $\tau\rightarrow \tau_c$ as\cite{WangPRE13,WangPoP13}
\begin{equation}
\partial_t F\left(\bm Z,t\right)+\left\{ F, H_0\right\}=Lim_{\tau\rightarrow \tau_c}\frac{1}{\tau}\left\langle \mathcal C_{\int}\right\rangle F= \partial_{\bm Z}\cdot\left[\frac{\left\langle\bm G(\tau_c)\right\rangle}{\tau_c}+\frac{\left\langle\bm G(\tau_c) \bm G(\tau_c)\right\rangle}{2\tau_c} \cdot \partial_{\bm Z} \right]F, \label{eq:lte}
\end{equation}
which is consistent with the Kolmogorov condition\cite{ZaslavskyPR02} $\left\langle\bm G ^{2n+2}\right\rangle=0,n\geq1$. Note that the FP operator contains the effects of nonlinear turbulence scattering.

It should be pointed out that the large-scale coherent structures, with the typical wave-length $\lambda_l \gg \lambda_c$, survive the ensemble-average. Eq. (\ref{eq:nlt}) and Eq. (\ref{eq:lte}) contain the meso-scale kinetics, such as the low-frequency zonal flows \cite{LinScience98} and the quasilinear transport induced by the growth of large-scale coherent modes \cite{KominisPRL10,WangPoP12}.

To deal with the macro-kinetics of the quasilinear transport in a confinement system, we introduce the macro-average operator $\left\langle\right\rangle_{\bm {\theta}}$, which can be taken as time-averaging over the periodic variables. For example, in a tokamak system, we can take $\bm Z =\left(J^1,J^2,J^3,\theta ^1,\theta ^2,\theta ^3\right)$, with $\theta ^1$ the gyro-angle, $\theta ^2$ the poloidal angle, $\theta ^3$ the toroidal angle, and $J^i$ the three independent constants of unperturbed motion.
\begin{equation}
\left\langle A\right\rangle_{\bm {\theta}}=\oint \mathcal J d \bm {\theta}A/\oint \mathcal J d \bm {\theta}.\label{eq:ma}
\end{equation}
Usually, one of the constants of motion ($J^1$), e.g. the toroidal canonical angular momentum, is essentially the generalized minor radius, and the other two can be chosen as the magnetic moment and the energy. In this case, $\bar F \left(\bm J\right)\equiv \left\langle  F\left(\bm J,\bm {\theta}\right)\right\rangle_{\bm {\theta}}$ and $F\left(\bm J,\bm {\theta}\right)=\bar F \left(\bm J\right)+\delta F\left(\bm J,\bm {\theta}\right)$ should be understood\cite{KominisPRL10,WangPoP12}.
For simplicity, we ignore the stochasticity and focus on the effects of the large-scale coherent modes; Eq. (\ref{eq:lte}) is reduced to
\begin{equation}
\partial_t F\left(\bm Z,t\right)+\left\{F, H_0\right\}= \partial_{\bm Z}\cdot\left[\partial_t\left\langle \bm G\right\rangle F \right]. \label{eq:ltec}
\end{equation}

Substituting $F\left(\bm J,\bm {\theta}\right)=\bar F \left(\bm J\right)+\delta F\left(\bm J,\bm {\theta}\right)$, one finds the linear solution\cite{WangPoP12}
\begin{equation}
\delta F =\left\langle \bm G ^{\bm J}\right\rangle \cdot\partial_{\bm J} \bar F\left(\bm J\right).\label{eq:deltaf}
\end{equation}
Substituting the result into Eq. (\ref{eq:ltec}) and taking the macro-average, one finds the quasilinear transport equation\cite{KominisPRL10,WangPoP12},
\begin{equation}
\partial_t \bar F\left(\bm J,t\right)= \partial_{\bm J}\cdot\left[\frac{1}{2}\partial_t\left\langle\left\langle\bm G ^{\bm J}\right\rangle \left\langle\bm G  ^{\bm J}\right\rangle \right\rangle_{\bm {\theta}} \cdot \partial_{\bm J} \bar F \right]. \label{eq:qlt}
\end{equation}
This demonstrates that the large-scale kinetic theory, Eq. (\ref{eq:lte}), includes both the familiar quasilinear transport and the nonlinear stochastic transport. It is a multi-scale kinetic theory, which can treat the usual macro-scale transport and the meso-scale kinetics including the large-scale coherent structures. The two-scale average method used here is different from the previous ensemble average in dealing with the quasilinear transport due to the coherent modes\cite{KominisPRL10,WangPoP12}, which is essentially the macro-average used here. Therefore, Eq. (\ref{eq:lte}) is different from Refs. [\onlinecite{KominisPRL10,WangPoP12}]; it is an important generalization of Refs. [\onlinecite{WangPRE13,WangPoP13}]; note that the second order truncation in Eq. (\ref{eq:S*}) can be replaced by high order expansion in the new framework.

The two-scale average [Eqs.(\ref{eq:ea} and \ref{eq:ma})] interprets the nonlinear stochastic transport [Eq. (\ref{eq:lte})] as the effect of the micro-turbulence scattering and the quasilinear transport [Eqs. (\ref{eq:ltec} and \ref{eq:qlt})] as the effect of long-range correlation of the meso-scale coherent modes. Further discussions on the multi-scale kinetics shall be given in the last section.

\section{\label{sec5}FRACTIONAL TRANSPORT EQUATION}

For a general stochastic or Hamiltonian chaotic system, the Kolmogorov condition may be oversimplified\cite{ZaslavskyPR02}, and one has to use the nonlocal instead of the local transport equation. The complicated operators in Eq. (\ref{eq:nlt}), with $n\geq1$, which account for the nonlocal or multi-scale effects, have not been included in the standard FP equation.
Consider a simple one-dimensional random-walk model problem. Suppose that at the initial time, the particle density distribution is $N\left(x,t\right)_{t=0}=N_0\left[1+\epsilon cos\left(kx\right)\right]$, with $\epsilon \ll1$.
Suppose the particles randomly walk in the $x$ direction with a step-size $-G\left(\tau_c\right)$ within the time interval $\tau_c$, which is the standard random walk diffusion model\cite{ChandrasekharRMP43}.

For a more general case, we introduce the modified Gaussian-type probabilistic distribution of $G\left(t\right)$,
\begin{equation}
\left\langle A\left[G\right(t\left)\right]\right\rangle=\frac{1}{\sqrt{4\pi Dt^{\alpha}}} \int_{-\infty}^{\infty} dz exp\left[-\frac{\left(z-ut^{\alpha}\right)^2}{4Dt^{\alpha}}\right] A\left(z\right), \label{eq:gauss}
\end{equation}
where, $u\neq0$ stands for the dynamic friction, which will be omitted for mathematical simplicity in the following. $\alpha=1$ stands for the standard diffusion due to the random walk model; $1>\alpha>0$ stands for the sub-diffusion; $2>\alpha>1$ stands for the super-diffusion case.

The particle distribution at time $t$ found from the standard diffusion equation $\partial_t N\left(x,t\right)= \partial_x \left( D \partial_x N \right)$, with the standard diffusivity\cite{ChandrasekharRMP43} $D=G^2\left(\tau_c\right)/2\tau_c$, is
\begin{equation}
\Delta N\equiv N\left(x,t\right)-N_0=\epsilon N_0cos\left(kx\right)exp\left(-k^2 D t\right). \label{eq:lte1ds}
\end{equation}
However, from Eq. (\ref{eq:nlt}), one finds
\begin{equation}
N\left(x,t\right)-N\left(x,0\right)= \sum_{n=1}^\infty \frac{1}{\left(2n\right)!}\left\langle \left[G \left(t\right)\right]^{2n}\right\rangle\partial_{x}^{2n} N\left(x,0\right), \label{eq:nlte1d1d}
\end{equation}
Substituting $N\left(x,0\right)$ and Eq. (\ref{eq:gauss}), one finds
\begin{equation}
\Delta N=\epsilon N_0cos\left(kx\right)exp\left(-k^2 D t^{\alpha}\right).\label{eq:nlte1ds}
\end{equation}
The solution with $\alpha=1$ agrees with Eq. (\ref{eq:lte1ds}). This illustrates the Fourier transform method to solve the problem.
For an initial distribution of the Gaussian packet,
\begin{equation}
N\left(x,t\right)_{t=0}=N_0 \frac{1}{\sqrt{2\pi a^2}}exp\left[-\frac{\left(x-x_0\right)^2}{2a^2}\right],
\end{equation}
one finds
\begin{equation}
N\left(x,t\right)=N_0 \frac{1}{\sqrt{4\pi D t^{\alpha}+2\pi a^2}}exp\left[-\frac{\left(x-x_0\right)^2}{4D t^{\alpha}+2a^2}\right].
\end{equation}
This suggests a fractional transport equation\cite{ZaslavskyPR02},
\begin{equation}
\frac{\partial ^{\alpha} }{\partial t^{\alpha}}N \left(x,t\right)= \partial_x \left( D_{\alpha} \partial_x N \right), \label{eq:fracte}
\end{equation}
\begin{equation}
D_{\alpha}\equiv Lim\left({\tau\rightarrow\tau_c}\right)\left\langle\frac{ G\left(\tau\right)G\left(\tau\right) }{2\tau ^{\alpha}}\right\rangle. \label{eq:fracd}
\end{equation}
Therefore, we have demonstrated that Eq. (\ref{eq:nlt}) includes the fractional transport, in addition to the standard (diffusive) transport.
Note that the fractional transport theory predicts the decaying time $\sim a^{2/\alpha}$, with $a$ the system size; this is reminiscent of the tokamak confinement scaling law\cite{WessonBook97} $\tau_E\sim a^{2/\alpha}$, with $a$ the minor radius of the device and generally $\alpha \neq1$.

\section{\label{sec6}CONCLUSIONS AND DISCUSSIONS}
In conclusion, we have shown that the familiar quasilinear transport, the nonlinear stochastic transport, and even the fractional transport, can be systematically derived from Liouville's theorem. The phase-pace multi-scale transport equation based on the first-principle and the two-scale average method proposed in this paper indicate that the nonlinear stochastic transport is due to the effect of the micro-turbulence scattering and the familiar quasilinear transport is due to the effect of long-range correlation of the meso-scale coherent modes. The new theory provides an opportunity to relate the anomalous transport observed in magnetic fusion plasma turbulence to the fractional transport.

To discuss the implications of the multi-scale kinetics, we rewrite the ensemble averaged transport equation, Eq. (\ref{eq:lte}) and Eq. (\ref{eq:ea}),
\begin{equation}
\partial_t F\left(\bm Z,t\right)+\left\{ F, H_0\right\}= \partial_{\bm Z}\cdot\left[\frac{\left\langle\bm G(\tau_c)\right\rangle}{\tau_c}+\frac{\left\langle\bm G(\tau_c) \bm G(\tau_c)\right\rangle}{2\tau_c} \cdot \partial_{\bm Z} \right]F, \label{eq:lte1}
\end{equation}
\begin{equation}
\left\langle f\left(\bm Z\right)\right\rangle=Lim_{l\gg\Delta Z\geq \lambda_c} \frac{1}{\Delta \bm Z ^6}\int_{\Delta \bm Z ^6}d \bm Z ^6 f\left(\bm Z\right); \label{eq:ea1}
\end{equation}
and the macro-averaged transport equation for the quasilinear transport due to the large-scale coherent modes, Eq. (\ref{eq:qlt}) and Eq. (\ref{eq:ma}),

\begin{equation}
\partial_t \bar F\left(\bm J,t\right)= \partial_{\bm J}\cdot\left[\frac{1}{2}\partial_t\left\langle\left\langle\bm G ^{\bm J}\right\rangle \left\langle\bm G  ^{\bm J}\right\rangle \right\rangle_{\bm {\theta}} \cdot \partial_{\bm J} \bar F \right], \label{eq:qlt1}
\end{equation}
\begin{equation}
\left\langle A\right\rangle_{\bm {\theta}}=\oint \mathcal J d \bm {\theta}A/\oint \mathcal J d \bm {\theta},\label{eq:ma1}
\end{equation}
with the linear behavior of the coherent modes given by $F\left(\bm J,\bm {\theta}\right)=\bar F \left(\bm J\right)+\delta F\left(\bm J,\bm {\theta}\right)$, and
\begin{equation}
\delta F =\left\langle \bm G ^{\bm J}\right\rangle \cdot\partial_{\bm J} \bar F\left(\bm J\right).\label{eq:deltaf1}
\end{equation}

The nonlinear stochastic transport [Eq. (\ref{eq:lte1})] is due to the effect of the micro-turbulence scattering while the quasilinear transport [Eq. \ref{eq:qlt1})] is due to the effect of long-range correlation of the large-scale coherent modes. In a general turbulent tokamak plasma, anomalous transport are composed of both the nonlinear stochastic transport due to the micro-scale scattering and the quasilinear transport due to the growing of the large-scale coherent modes.

Zonal flows \cite{LinScience98}, as the meso-scale modes, do not directly generate transport by themselves, due to their toroidal symmetry. However, they can regulate the nonlinear stochastic transport by taking energy from the background turbulence and reducing the radial correlation length of the turbulence. This is the well-known paradigm to confinement improvement.

The present theory implies a new approach to confinement improvement. The scenario is briefly summarized as follows.
The development of the linearly unstable modes nonlinearly generate micro-scale stochastic turbulence, meanwhile they can also nonlinearly generate large-scale modes that are linearly stable, for example, the zonal flows. These coherent modes, either the linearly unstable or the linearly stable modes except the toroidal symmetric zonal flows, when they are growing, they contribute to the quasilinear transport [see, Eq. (\ref{eq:qlt1})] and Eq. (\ref{eq:deltaf1}). However, in a steady-state turbulence these coherent modes cease to contribute to the quasilinear transport, since they cease to grow due to the nonlinear stochastic damping\cite{WangPRE13,WangPoP13}. Therefore, when the turbulence energy moves from the micro-scale stochastic fluctuation to the large-scale coherent modes and eventually reach a new steady-state, the anomalous transport is reduced and the confinement is improved. The point is that all the steady-state large-scale coherent modes, including the large-scale symmetric zonal flows, do not contribute to the anomalous transport by themselves. In this sense the nonlinearly stable large-scale coherent modes can be regarded as the nonlinearly self-organized stable large-scale structures.

\begin{acknowledgments}
This work was supported by the National Natural Science Foundation of China under Grant No. 11175178, No. 11375196 and the National ITER program of China under Contract No. 2014GB113000.
\end{acknowledgments}

\nocite{*}


%

\end{document}